# Surveying *Wonderland* for many more literature visualization techniques

Richard Brath*

Uncharted Software Inc.

**ABSTRACT**

There are still many potential literature visualizations to be discovered. By focusing on a single text, the author surveys many existing visualizations across research domains, in the wild, and creates new visualizations. 58 techniques are indicated, suggesting a wider variety of visualizations beyond research disciplines.

**Index Terms**: H.5.2 [User Interfaces]: Screen design; I.2.7 [NLP]: Text analysis; I.7.m [Text Processing]: Miscellaneous; I.5.4 [Pattern Recognition]: Text processing; J.5 [Arts and Humanities]: Literature.

## 1 INTRODUCTION

Are current data visualization techniques too narrow for digital humanities? There may be many more literature visualizations in-the-wild than in the visualization research community, and these can be used to inform more visualizations.

The foundations of data visualizations originated from quantitative and categoric data analysis in statistics [1,2], cartography [3] and computer science [4]. There are 400+ visualizations for textual analysis from research [5-8]—but, many extend quantitative visualizations, such as more than 100 variants of graphs. For example, there tend to be few visualizations of letters or full text of paragraphs. Current text visualizations are too constrained. We are in danger of McLuhan's dictum: "We shape our tools, thereafter our tools shape us." [9,10]

Broader visualizations enable wider analysis. Many researchers advocate design alternatives. Munzer says "A fundamental principle of design is to consider multiple alternatives then choose the best." [11] Roberts et al promote the *Five Sheet Design Method* to foster divergent thinking and facilitate design alternatives [12]. Hinrichs et al argue for novel experimental visualizations to provoke insights, re-interpret knowledge and mediate ideas [13]. Buxton shows the need to ideate through design alternatives [14]. Tohidi indicates multiple designs facilitate design criticisms and help identify design problems [15].

* rbrath@unchartedsoftware.com

Bertin [3] created 90 different visualizations of a single dataset indicating the breadth of quantitative visualization. Symanzik et al[16] found 24 different visualizations of Titanic data. Expanding on this approach (and [17]): Can a single literary dataset aid finding a breadth of text visualizations?

This paper focuses on one fictional book: Lewis Carroll's *Alice's Adventures in Wonderland,* as it has been in continuous publication for 150 years, is widely known, is available digitally, and is used in visualizations both inside and outside the visualization research community. The following survey uncovers dozens of visualizations, which, in turn, frames new visualizations, forming an overall contribution of 50+ literature visualization techniques.

## 2 SURVEY OF WONDERLAND VISUALIZATIONS

Using text search, image search and scholar search with the terms *visualization, diagram* or *infographic* and variants of the book title returned 41 visualizations with some common themes:

**A. Visualization and Humanities Research.** The text of *Alice* is used in now common techniques such as Figure **1)** wordtrees [18]; **2**) tag clouds [19]; **3**) graphs [20,21]; and **4**) storylines [22]. It has also been used in the wonderfully experimental TextArc (**5**) [23]. Many of these visualizations work at the unit of or a word or two at a time. Word size is used to indicate frequency. Associations between words is indicated with lines. Wordtrees extends frequency to sentences. In Storylines, lines encode sequence and character proximity.

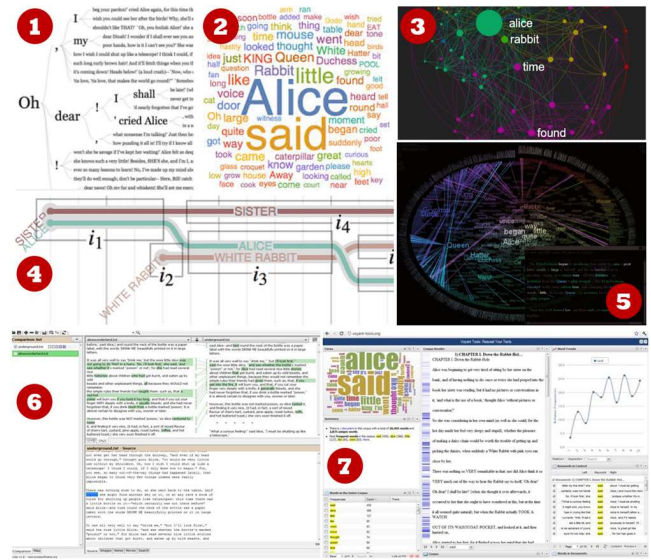

Text analysis systems incorporate visualization components with linked interactions, such as Figs. **6**) Juxta [24], and **7**) Voyant [25], with screenshots of *Alice*. In addition to word analysis, these systems include full text with keyword markup. Interactions, such as linking and filtering, enable the viewer to narrow in and compare original context. Language departments use text analytics and visuals may accompany publications such as Figs. **8**) word concordance aligning and bolding modifiers for skimming; **9**) noun topics and classification sources indicated via caps and italics; and **10**) location of topic words within the document [all 26].

**B. *Alice* in NLP.** In Natural Language Processing (NLP) *Alice* has been used for low-level syntax and grammar such as **11**) parse trees [27]; and **12**) other hierarchical displays [28]. Visualizations of terms, topics and relations include **13**) graph of terms and related associations; **14**) extracted topic words per chapter [both29]; **15**) plots of extracted social network metrics per character [30]; **16**) lines indicating topic recurrence (persistent homology) [31]; **17**) a chart of character frequency [32] and **18**) a chart of emotions [33]. These visualizations are utilitarian diagnostics of word statistics and modelled associations with simple representations.

**C. *Alice* in Fine Arts and Computer Graphics.** Academics in arts and computer graphics create different layouts and structure, such as, **19**) the microstructure of phonemes across sentences and languages [34]; **20**) a tool for drawing non-linear layouts of text [35]; and **21**) digital micrography, wherein text flows within expressive shapes [36].

**D. *Alice* Visualized in the Wild.** There are many other visualizations beyond formal research and research tools. Some extract quantitative data, such as **22,23**) Alice's changing height [37,38], **24,25**) timelines [39,40], or **26**) instances of Alice's dress [41]. These use common visualization techniques such as line charts, timelines and small multiples.

More varied are visualizations related to characters and relationships, such as Figure **27**) a interactive *lap book* for teaching [42]; **28**) a pop-up book, summarizing *Alice* into six scenes [43]; **29**) itemized entities, e.g. color-coded characters, places and events [44]; **30**) infographics of themes, characters and relations and motifs [45]; **31,32**) graph diagrams showing key characters and relations [46,47]; and **33,34**) mind-maps of characters, setting, problems and the solution [48,49].

Plot and themes include examples following chapter sequence, such as **35**) a bar chart, one bar per chapter [50]; **36**) a wheel where each wedge is a chapter, extending by number of sections in the chapter, color coded by the theme(s) per section [51]; **37,38**) plot diagrams, summarizing steps leading to climax and resolution[52,53]; and **39**) a multivariate design with rings per chapter; bars indicating content measures of puns, logic, rhyme, etc., and color-coded character dots [54]. Some of these visualizations attempt to uncover structure –

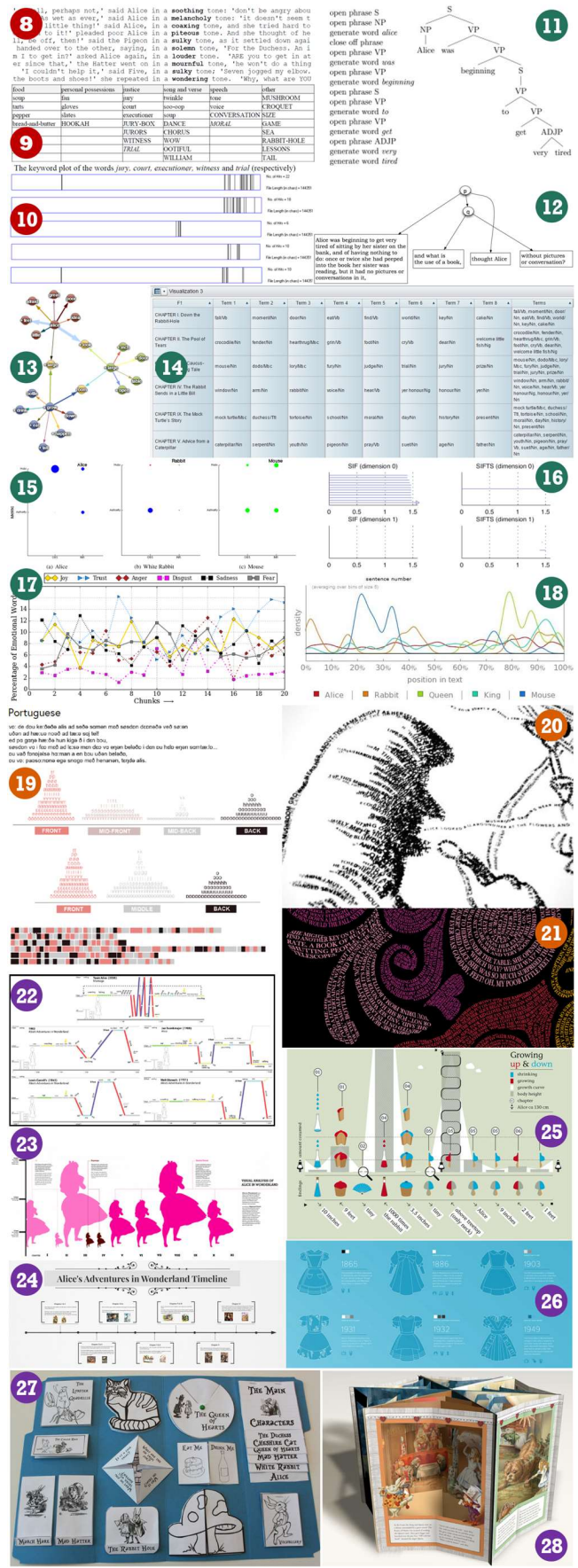

similar to Posavec's *Writing without Words* [55] or Cherny's extracted story arcs [56].

Designers have created unique layouts in limited edition books, such as **40**) variation in type size and layout [57]; and **41**) superimposed extracts of logical inversions [58]. The typographic manipulation creates a secondary reading across paragraphs and chapters while maintaining the original text.

**Summary.** These art and in-the-wild examples may be questionable (accurate? legitimate?), but they do use visualization cues such as color-coding of words or backgrounds; organizing groups of entities; and using connectors. Unlike prior visualizations, these examples tend to be text heavy ranging down to phonemes and up to abstractive descriptions of characters, extracted quotes, etc., as summarized in Table 1. Typographic enrichment, such as bold, italics, are also more common in the wild.

| Table 1: Analysis of prior *Alice* visualizations | | Number of visualizations by text use in plot area | | | | | | Visual encoding | | |
|---|---|---|---|---|---|---|---|---|---|---|
| Domain | Number of Visualizations | No words in plot area | Sub-word e.g. phonemes | Labels (1-2 words) | Abstractive phrases/ sentences | Extractive phrases/ sentences | Critical analysis | Typographic (bold, italics, etc) | Visual (size, color, etc) | Position only |
| Vis & Humanities | 10 | 1 | 0 | 6 | 0 | 3 | 0 | 2 | 7 | 1 |
| NLP | 8 | 4 | 0 | 3 | 0 | 1 | 0 | 0 | 4 | 4 |
| Art & CG | 3 | 0 | 1 | 0 | 0 | 2 | 0 | 2 | 2 | 0 |
| In the Wild | 20 | 3 | 0 | 4 | 9 | 2 | 2 | 8 | 12 | 5 |
| TOTAL | 41 | 8 | 1 | 13 | 9 | 8 | 2 | 12 | 25 | 10 |

## 3 ADDITIONAL VISUALIZATIONS

The prior 41 visualizations indicate many ways to visualize a text—are there more? Other novel visualization techniques exist which have not covered *Alice.* E.g., parallel tag clouds [59] and self-organizing maps [60], but largely focus on words and associations by proximity or explicit linking. Tendril [61] creates a 3D structure based on hyperlinks for an interactive branching reading. But few research visualizations are similar to Figures 19-41.

To consider a wider range of visualizations for literature is to reconsider what can be visualized and how it can be visualized, i.e. a design space exploration. A design space is the set of all the parameters for design variation—in visualization this includes data such as categories, quantities and literal text; visual attributes such as color, size, bold and italic; scope ranging individual letters to chapters; and various layouts—as suggested by Table 1. Many authors have defined the design space [3-4,62-68], the maximal set of design space definitions is considered here. Using Table 1's outline of text in the plot area, consider additional visualizations of *Alice*:

**A. Sub-words.** Figure 19 (and prior work e.g. [69]) indicates potential for sub-word visualization. Figure **42** shows neologisms, words deliberately misspelled by Carroll, either lengthened with added letters (shown with high x-height) or shortened (removed letters with low x-height). **43** shows a song from *Alice*, with a shifting baseline to indicate note pitch, letter-width to indicate note duration, red text to indicate rhyme phoneme, and yellow box to indicate Carroll's edits of the original. Sub-word visualization is under-explored: it could be used to indicate stylistic devices such as alliteration, assonance, consonance, cacophony, etc.

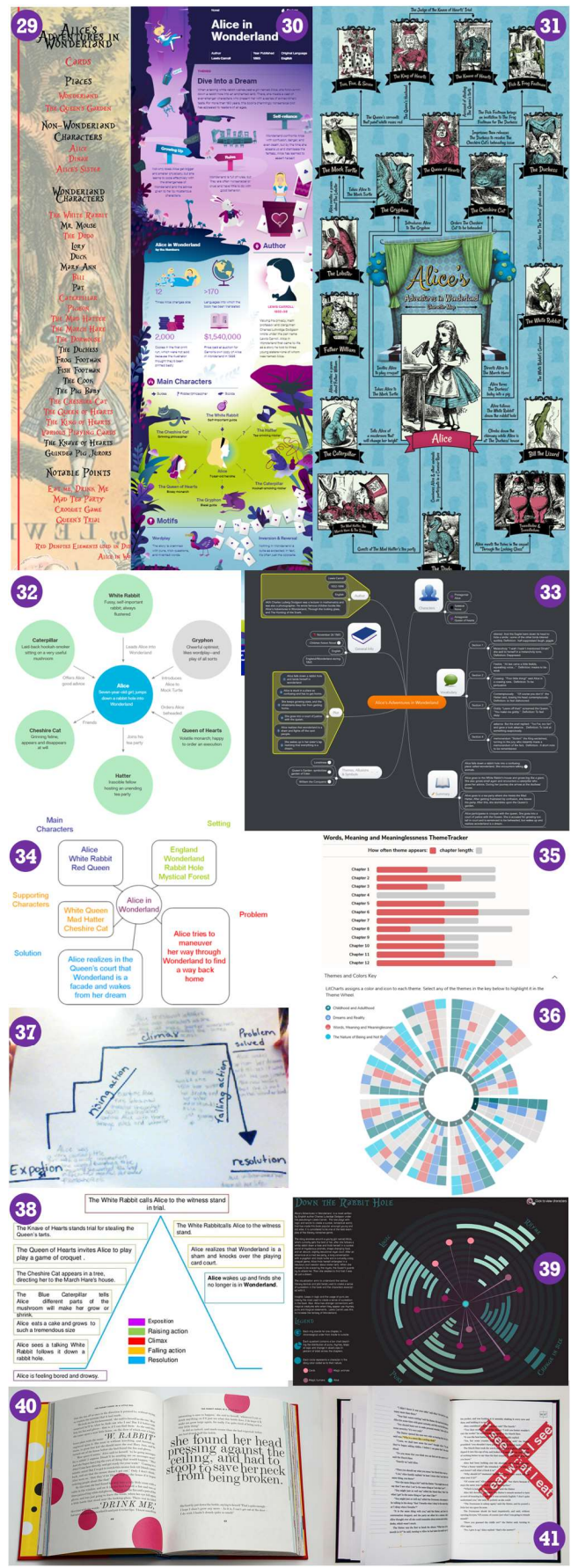

**B. Words and Word Pairs**. Many text visualization techniques extract words then visualize (e.g Figs. 2-5,9-10). These words may be used as literal markers in a visualization such as **44,** depicting emotions per character, based on an NLP emotion lexicon [70]. Visualizations that show all labels (Figs 5,9,43,44) do not require slow interaction to reveal labels. Instead, the labels can be read quickly with a shift in attention, yielding faster scanning [68] and potential for serendipitous discovery [71].

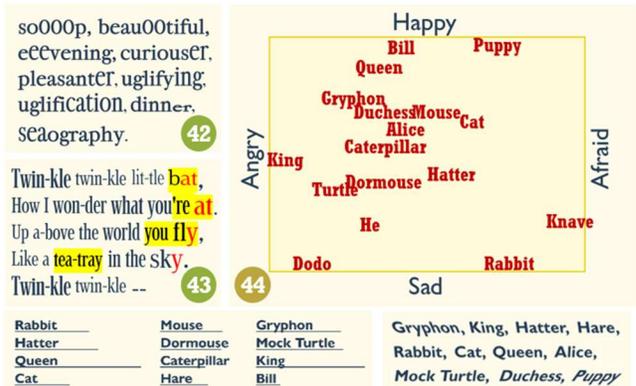

Labels can encode information via font, color etc., such as Fig. 9, or: **45**) underline length to indicate word frequency—a more compact, more perceptually accurate representation than a word cloud; **46**) oblique angle to indicate character sentiment (extracted via NLP), with the most negative character sloping steep left (Gryphon) to most positive sloping steep right (Puppy)[17], and **47**) characters over time with character type indicated by typeface (e.g. birds and cards tend to occur together). **48** shows the most common bigrams in *Alice*: left is the first word in the pair, the following list of words are the most frequent following words, with font weight indicating frequency of occurrence of the pair, e.g., "oh dear" is more frequent than "oh my", which is more frequent than "oh how". Pair analysis can do more, e.g. nouns and nearby modifiers [72]. **49** is a textual stem-and-leaf plot of characters and adverbs (Alice is timid, the Queen is furious, the Mouse is cross). Formats can be combined across many variables and cross-referenced. For example, **50** shows four metrics per character: note how the birds (Lory, Duck, Dodo, Eaglet) recede – not highly ranked by any metric. The birds aren't talkative, but speaking doesn't correlate to popularity: the CHESHIRE CAT is near the top of most lists (wide, dark, all caps).

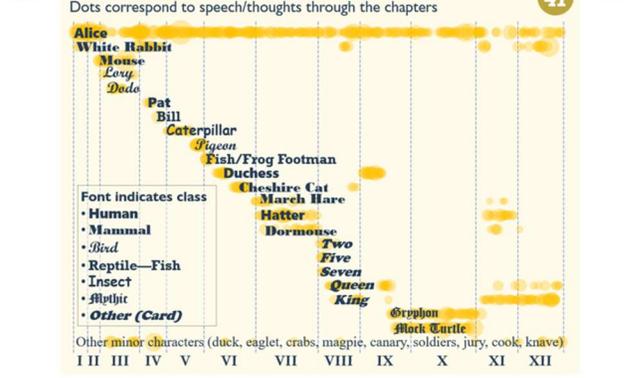

The depiction of NLP analytics, such as amount spoken, emotion or sentiment, allows a reader to compare to their personal knowledge, e.g., NLP in Fig. 46 has modeled the Queen as neutral whereas the reader may perceive the Queen to be angry. Layering in extra data via formats invites critical analysis of both text and algorithmic bias.

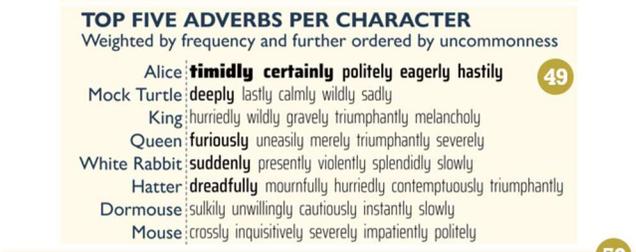

**C. Sentences and Paragraphs**. Many rhetorical devices are associated with sentence construction and repetition of phrases across sentences. Fig. **51** is a syntax diagram [73], similar to a word tree plus loops for repetition. Font weight indicates frequency of successive phrases. A railway diagram can represent actual text or generative text (e.g. auto-completion or transformer models). This diagram indicates text that Carroll wrote, e.g., "will you walk a little faster"; or used to construct phrases that Carroll did not write, e.g., "will you, won't you, will you walk a little faster." **52** is a kelp fusion underlay [74], highlighting and linking repetitions of two or more words, similar to Poemage [75]. Beyond logical inversions in this dialogue are many more repetitions, such as "the same thing" and "you might just as well say," verbally aligning the mad partiers against Alice.

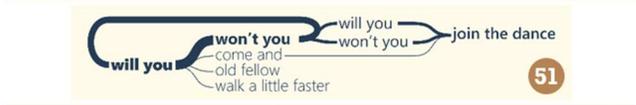

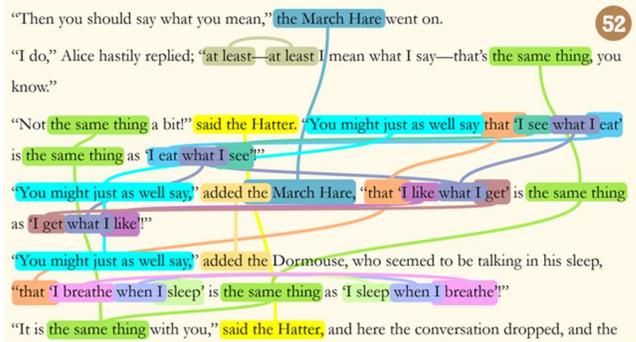

With prose, layout is arbitrary. Interactive layouts can be adjusted, for example, to align repetitions. Unlike a word tree or keywords-in-context (KWIC) [76]), the full linear sequence of text remains, and can align words defined by the user (**53**) or automated, e.g. key nouns (**54**).

Fig. **55** shows NLP statistics and annotation counts [77] per chapter. Each chapter and a portion of opening text are shown on each line. Stats are shown as colored bars, bold or underline: e.g. *Pig and Pepper* is the most disgusting chapter.

Fig. **56** is an adjacency matrix indicating dialogue from one character (vertically) to another

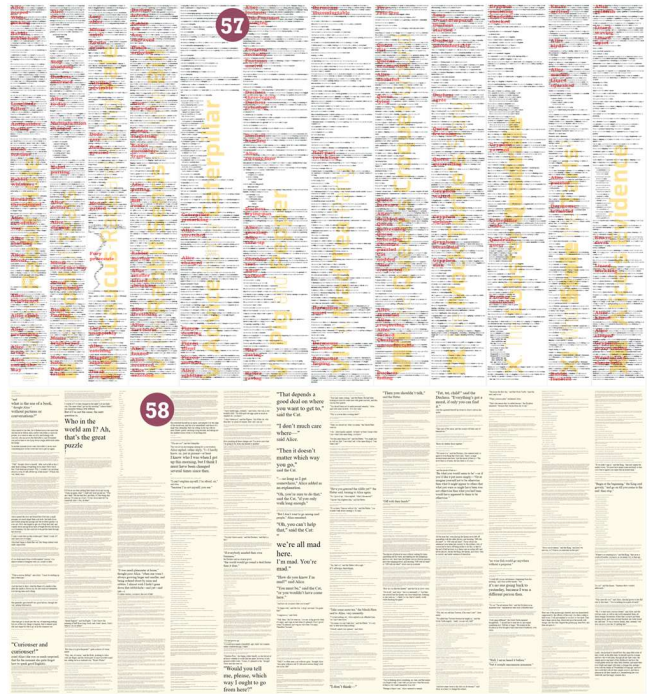

(horizontally). Each cell contains the dialogue [17]. Cells are expanded if the next cell to the right would otherwise be empty, and some cells are truncated. Colored content indicates repeated word sets: the Duchess repeats "moral of that is"; the Queen says variants of "off with his/her/their head".

**D. Chapters and Document**. Document navigation can be aided with enhanced chapters, paragraphs, or lines [78]. With posters, 4k displays, (or blankets [79]), the entire text of *Alice* can be displayed, e.g. Fig. **57**. For sizable paragraphs, a noun and verb are extracted, enlarged and set in red (e.g. "Alice peeped" or "Mouse patted"). Chapter titles are presented even larger behind the text. **58** shows the text of *Alice*, with font size successively increased for 173 popular *Alice* quotations cited on the Internet, making sentences stand-out such as *Who in the world am I? Ah, that's the great puzzle*, or, *We're all mad here*. [80-84]. These manipulations create textual landmarks, which aid quick navigation to the text of interest, and then the detailed text attended to; or provide alternative readings with context.

## 4   DISCUSSION

These 58 examples, from research, in the wild and constructed, visualize *Alice* many ways. So what?

Common visualizations such as line charts, bar charts, graphs and scatterplots can be extended further, e.g. prose text, formats, alternate layouts. Marks can range beyond words down to subwords, sentences, paragraphs and documents. The breadth allows the text to be deconstructed and analyzed e.g. for neologisms, prosody, emotion, sentiment, timelines, character types, character popularity, character descriptions, catch phrases, key nouns, logical fallacies, summarization, landmark extraction, and the potential for much more.

Reviewing these with seven visualization designers (advanced degrees and industry experience in visualization, interaction design, media design, and graphic design), included responses: "I had no idea the range of possible solutions could be so broad," "It's interesting what you can do when you cast a wider net," and, "I really like how there are opportunities to visually manipulate text more than words." There was also discussion regarding usability with consensus that interaction can aid use.

Some examples split text, manipulate layout and superimpose text, creating different readings, and interrupting linear reading. This is countered that post-modernists create multiple readings though type and layout manipulation, as seen in periodicals such as *Ray Gun*, *Octavo*, and *Emigre*.

## 5 CONCLUSION

Most important, this review and the constructions shows there are many different ways to visualize a text, beyond existing visualizations. Each technique extracts, highlights or modifies aspects of the text for different insights. The intent of many permutations and analytics illustrates the potential for more varied analyses and potential novel insights.

## ACKNOWLEDGEMENTS

The author's images are open source, licensed under CC BY SA 4.0 and available on the Internet.